\documentclass[aps,prl,twocolumn,superscriptaddress,footinbib,amsmath,amssymb]{revtex4}
\usepackage{blindtext}
\usepackage{gensymb}
\usepackage{dcolumn}
\usepackage{bm}
\usepackage[pdftex]{graphicx,color} 
\usepackage{epstopdf} 
\usepackage{graphicx}
\usepackage{float}
\usepackage[english]{babel}
\usepackage[normalem]{ulem} 
\begin{document}
\title{Looking inside three-dimensional (3D) silicon photonic band gap crystals}

\author{W. L. Vos}
\email{w.l.vos@utwente.nl}
\affiliation{Complex Photonic Systems (COPS), 
MESA+ Institute for Nanotechnology, 
University of Twente, P.O. Box 217, 7500 AE Enschede, The Netherlands}
\homepage{www.photonicbandgaps.com}

\author{D. A. Grishina}
\affiliation{Complex Photonic Systems (COPS), 
MESA+ Institute for Nanotechnology, 
University of Twente, P.O. Box 217, 7500 AE Enschede, The Netherlands}

\author{P. Cloetens}
\affiliation{European Synchrotron Radiation Facility (ESRF), 
B.P. 220, 38043 Grenoble C{\'e}dex, France}

\author{C. A. M. Harteveld}
\affiliation{Complex Photonic Systems (COPS), 
MESA+ Institute for Nanotechnology, 
University of Twente, P.O. Box 217, 7500 AE Enschede, The Netherlands}

\author{P. W. H. Pinkse}
\affiliation{Complex Photonic Systems (COPS), 
MESA+ Institute for Nanotechnology, 
University of Twente, P.O. Box 217, 7500 AE Enschede, The Netherlands}

\date{Paper at the Symposium on occasion of prof. Ad Lagendijk's birthday, version 2016-09-08.}

\begin{abstract}
We have performed an x-ray holotomography study of a three-dimensional (3D) photonic band gap crystal.
The crystals was made from silicon by CMOS-compatible methods. 
We manage to obtain the 3D material density throughout the fabricated crystal. 
We observe that the structural design is for most aspects well-realized by the fabricated nanostructure. 
One peculiar feature is a slight shear-distortion of the cubic crystal structure. 
We conclude that 3D X-ray tomography has great potential to solve many future questions on optical metamaterials. 
\end{abstract}

\maketitle

\section{Introduction}
\begin{figure}[tbp]
\includegraphics[width=1.0\columnwidth]{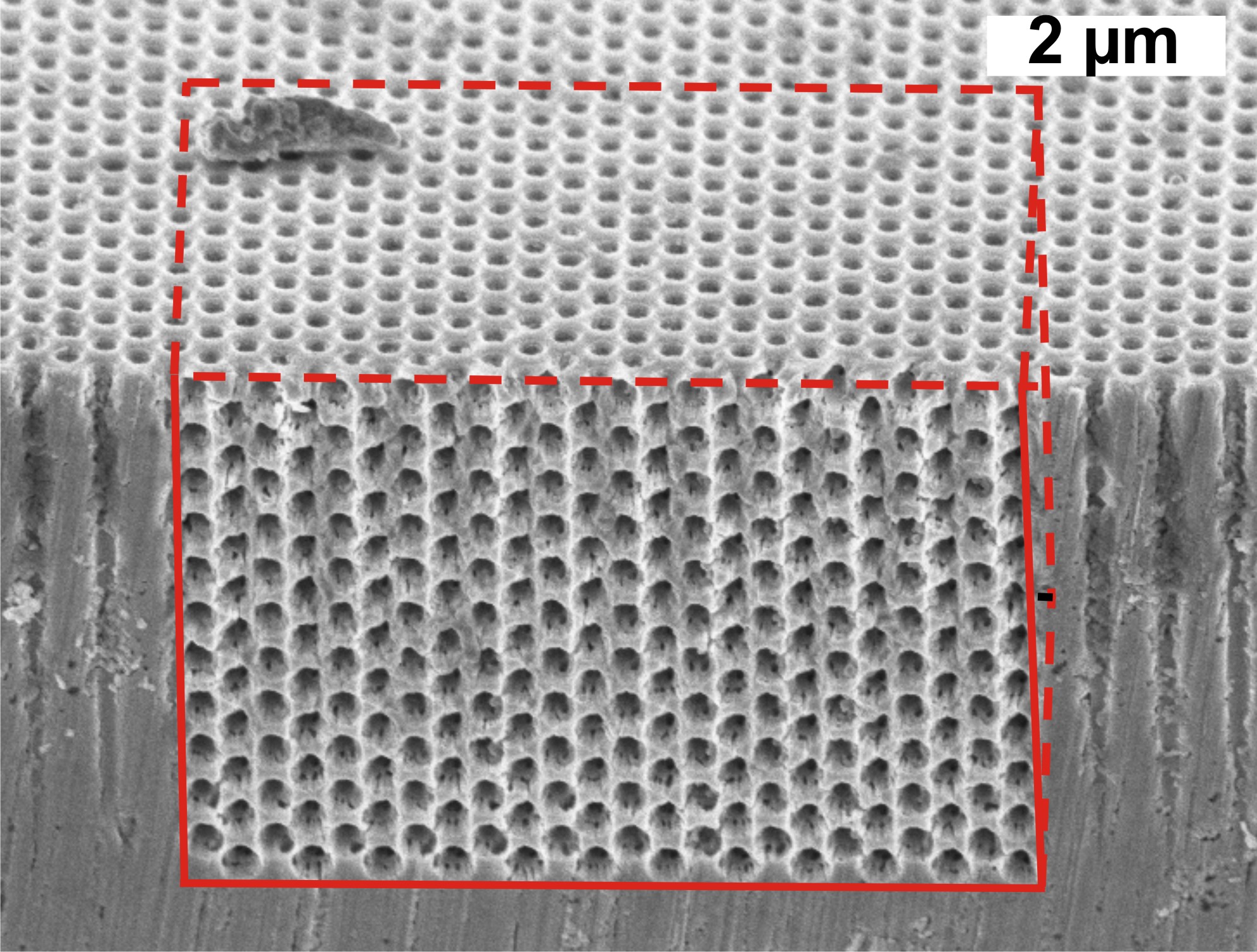}
\caption{
Scanning electron microgaph (SEM) of the surface of a 3D Si inverse woodpile photonic crystal. 
The \textit{estimated} extent is shown by the dashed lines. 
The 3D crystal is surrounded by a large 2D array of pores that are first etched in the Z-direction.
Image from Ref.~\cite{Leistikow2011PRL}. }
\label{fig:SEMstructure} 
\end{figure}

In nanofabrication, it is an ongoing challenge to characterize the structure of a real sample, in particular of three-dimensional (3D) materials~\cite{Ergin2010Science,vandenBroek2012AFM}. 
This challenge holds notably for nanophotonic metamaterials whose properties and functionality are determined by their complex structure with feature sizes $d$ comparable to, or even smaller than the wavelength of light: $d \leq \lambda$~\cite{Novotny2006book,Joannopoulos2008book}. 
These classes of nanomaterials are from necessarily opaque, hence current high-resolution optical microscopy techniques have insufficient penetration depth, apart from the usual resolution limitations~\cite{Hell2007Science}. 
Similarly, while scanning electron microscopy offers a very high spatial resolution, this method has such a small penetration depth that only the surface of a sample can be studied, as shown in the example in Figure~\ref{fig:SEMstructure}. 

In the field of nanophotonics and metamaterials, X-ray techniques have received surprisingly little attention, which is surprising, given that these techniques are ideal to characterize the 3D structure of opaque. 
Notably, X-rays have excellent penetration depth, the methods are generally non-destructive and readily provide nanometer-scale spatial resolution. 
Here, we study a Si photonic band gap crystal with a 3D diamond-like structure by X-ray holotomography~\cite{Cloetens1999APL}. 
The crystal has a so-called inverse woodpile structure~\cite{Ho1994SSC}. 
Similar crystals reveal a strong stabilization of the excited state of embedded semiconductor quantum dots, as demonstrated by  ten-fold spontaneous emission inhibition of many emitters simultaneously~\cite{Leistikow2011PRL}. 

\section{Samples and methods}
The material density distribution of the inverse woodpile crystals is defined by two perpendicular 2D arrays of pores\cite{Ho1994SSC}, as shown in the realized crystal in Figure~\ref{fig:SEMstructure}~\cite{Leistikow2011PRL,vandenBroek2012AFM}. 
In this report, the axis are defined such that pores are running the in the $Z$ and $X$-directions.  
Both 2D arrays have a centred rectangular structure with lattice parameters $(a,c)$, that correspond to diamond $hkl = 110$ faces.
For cubic diamond-like crystal structures, the lattice parameters are hence designed so as to fulfill the condition $a/c=\sqrt2$. 

Holotomography experiments were performed at the European Synchrotron Radiation Facility (ESRF), beamline ID-16NI, in the course of experimental run HC-2520. 
In brief, the X-ray beam with $17$ keV photon energy is focussed upstream from the sample. 
By varying sample-to-detector-distance we modify the diffraction pattern, as we operate in the Fresnel regime. 
To obtain data at all spatial frequencies, we collect data at multiple distances. 
At each distance, $1500$ images were recorded while rotating the sample from $0\degree$ to $180\degree$. 
Phase maps are obtained from the intensities at every angle, followed by standard tomographic reconstruction based on the inverse Radon transform to obtain the electron density distribution. 

\begin{figure}[tbp]
\includegraphics[width=1.0\columnwidth]{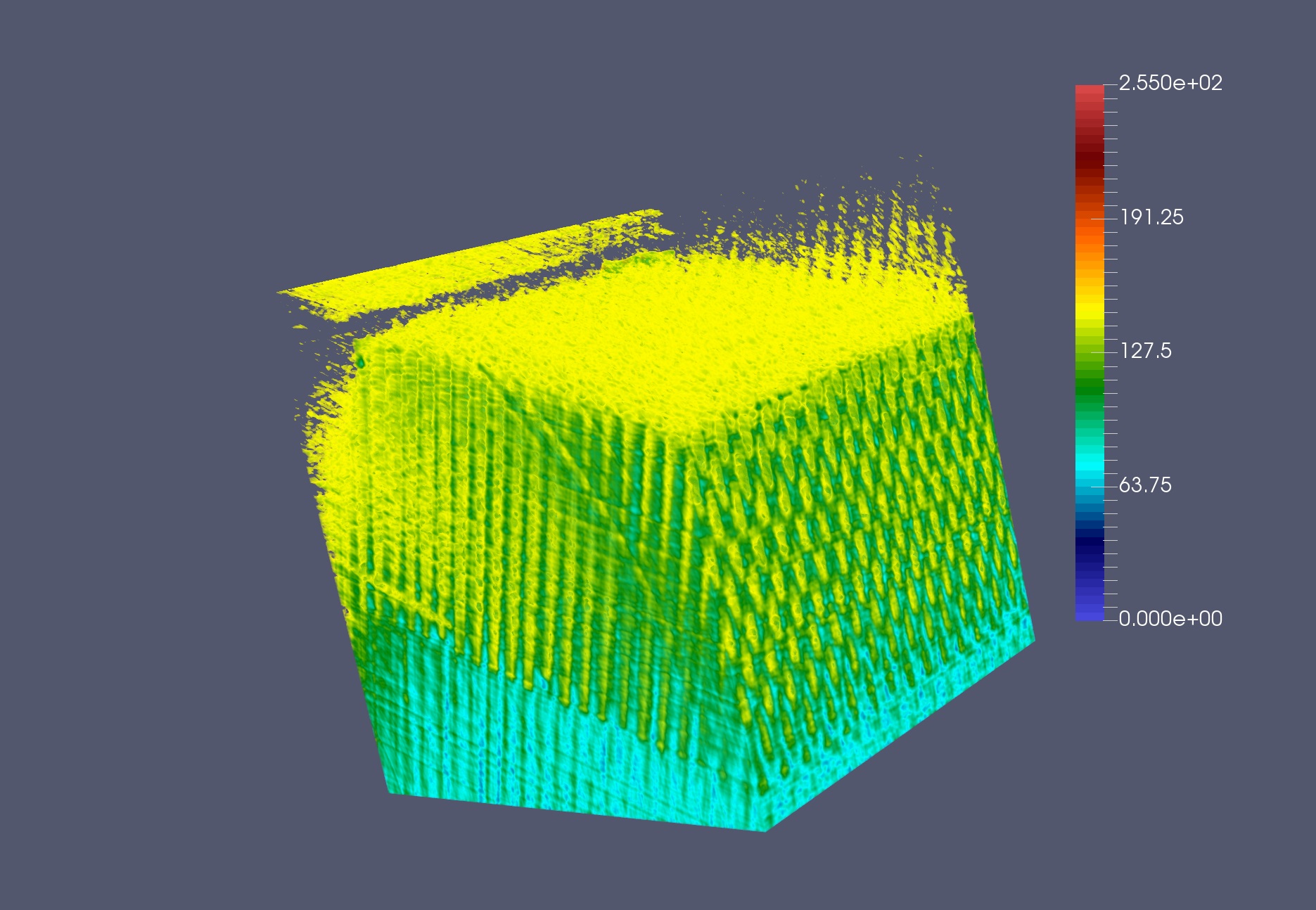}
\caption{\label{fig:structure} 
Bird's-eye view of the reconstructed sample volume. 
The color scale is the material density $\rho (X,Y,Z)$ interpolated between air and Si, the latter set to 255.  }
\label{fig:tomo-volume}
\end{figure}

\section{Results and discussion}
The alignment of the pores is a crucial step in the nanofabrication that can only be assessed \textit{in situ} by X-ray tomography. 
Moreover, it is crucial to verify the cylindrical pore geometry, and that pore depth that ultimately limits the crystal size. 

Figure~\ref{fig:tomo-volume} shows the reconstructed volume of one crystal.
A closer inspection~\cite{Grishina2016tbp} indeed reveals two sets of pores running in the $Z$ and the $X$-directions, matching the design. 
It appears that the angle between the pore arrays systematically deviates by a few degrees from the $90^{o}$ design. 
This means that the crystal structure is not truly cubic but monoclinic. 
Band structure calculations~\cite{Woldering2009JAP} reveal that the 3D photonic band gap remains robust under this effective shear deformation: the photonic band gap becomes hardly narrower, changing from $24 \%$ relative bandwidth to more than $21 \%$, which is plenty for strong spontaneous emission control~\cite{Leistikow2011PRL}. 

\begin{figure}[tbp]
\includegraphics[width=1.0\columnwidth]{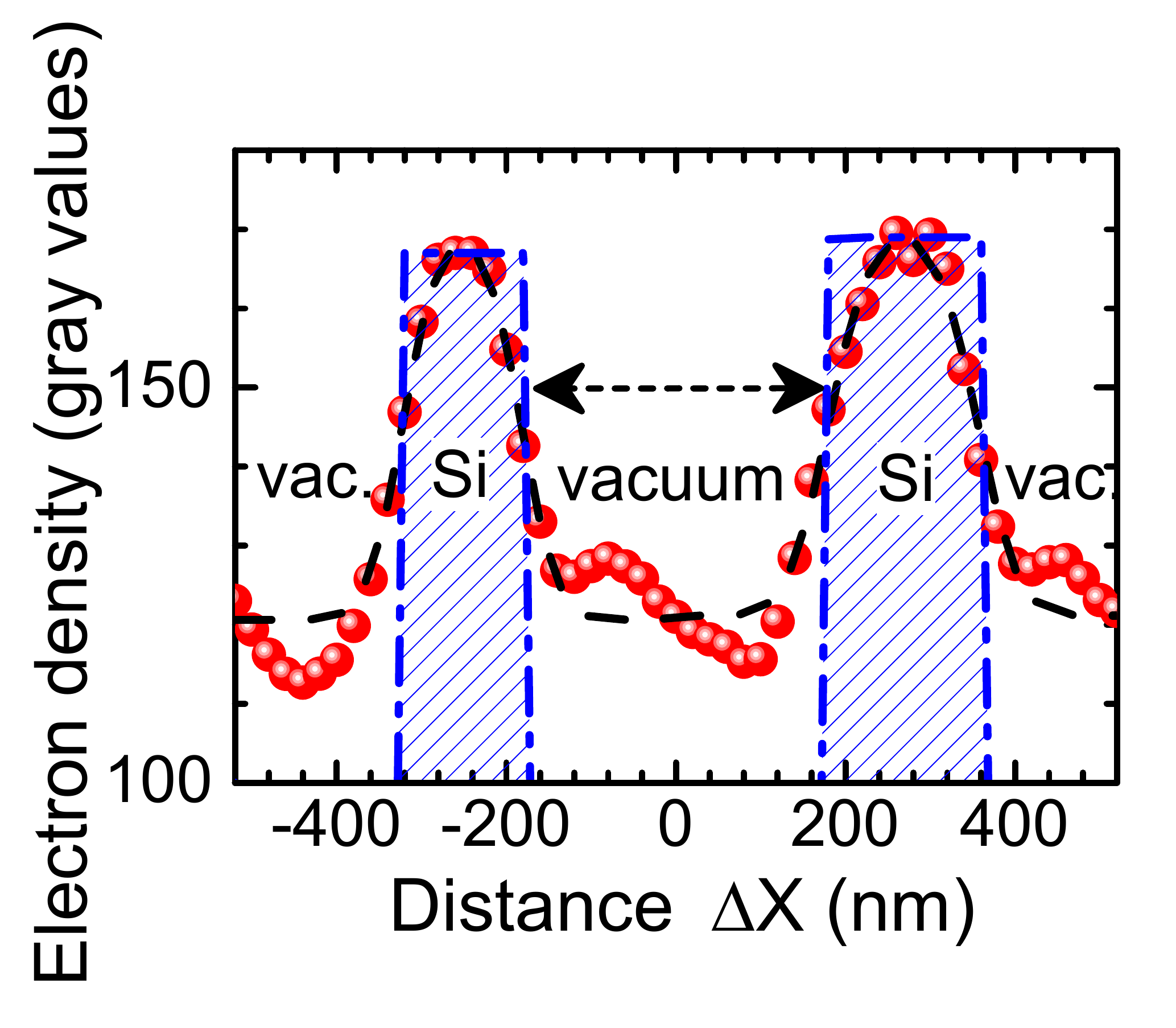}
\caption{\label{fig:X-cross-section} 
Density profile in the $X$-direction through pores in Si (red circles).
The black dashed curves is the convolution of a square density function (blue hatched bars) and a Gaussian resolution function.}
\end{figure}
Figure~\ref{fig:X-cross-section} shows a cross-section in the $X$-direction through three pores in the reconstructed volume. 
To interpret the density, we propose a simple model of a binary normalized density function for silicon ($\rho = 1$) and vacuum ($\rho = 0$) convoluted with a Gaussian resolution function.
The adjustable model parameters are the width $w$ of a Si-wall, the Gaussian resolution $\sigma$, the amplitude, and a background.
The data are well modeled with two Si-walls with widths $w_1 = 140$~nm and $w_2 = 176$~nm. 
Given the $20$ nm voxel size and the spatial resolution, this agrees with the design $w = 156$~nm, hence we conclude that the structure was faithfully realized. 
The resolution widths ($\sigma_1 = 32$~nm and $\sigma_2 = 37$~nm) agree well with the measured X-ray beam size. 
Modelling of the structure offers a straightforward future path to input tomographic structural data into $ab-initio$ numerical codes (FEM, FDTD) to compute nanophotonic properties. 
This opens the prospect to do \emph{model-free} calculations of metamaterial to predict experiments and device functionality. 


\section{Summary and outlook}
We have performed an x-ray holotomography study of a three-dimensional (3D) photonic band gap crystal.
The crystals was made from silicon by CMOS-compatible methods. 
We manage to obtain the 3D material density throughout the fabricated crystal. 
We observe that the structural design is in most aspects well-realized by the fabricated nanostructure. 
One exception is a slight shear-distortion of the cubic crystal structure. 
We thus conclude that 3D X-ray tomography has great potential to solve many future questions on optical metamaterials for much nanophotonic research and applications, including cavity arrays, physically unclonable functions, and precise localization of light emitters for enhanced efficiency.  

\begin{thebibliography}{}

\bibitem{Cloetens1999APL}
P. Cloetens, W. Ludwig, J. Baruchel, D. van Dyck, J. van Landuyt, J.P. Guigay, and M. Schlenker, 
\emph{Holotomography: Quantative phase tomography with micrometer resolution using hard synchrotron radiation x rays},
Appl. Phys. Lett. {\bf 75}, 2912 (1999)

\bibitem{Ergin2010Science}
T. Ergin, N. Stenger, P. Brenner, J.B. Pendry, and M. Wegener, \emph{Three-dimensional invisibility cloak at optical wavelengths}, 
Science {\bf 328}, 337 (2010)

\bibitem{Grishina2016tbp}
D.A. Grishina, P. Cloetens, C.A.M. Harteveld, and W.L. Vos,
to be published (2016) 

\bibitem{Hell2007Science} 
S.W. Hell, 
Science {\bf 316}, 1153 (2007).

\bibitem{Ho1994SSC}
K. M. Ho, C. T. Chan, C. M. Soukoulis, R. Biswas, and M. Sigalas,
\emph{Photonic band gaps in three dimensions: new layer-by-layer periodic structures}, 
Solid State Commun. \textbf{89}, 413 (1994).

\bibitem{Joannopoulos2008book}
J.D. Joannopoulos, S.G. Johnson, J.N. Winn, and R.D. Meade,
\emph{Photonic crystals, Molding the flow of light}
(Princeton University Press, Princeton NJ, 2008) 2nd Ed.

\bibitem{Leistikow2011PRL}
M. D. Leistikow, A. P. Mosk, E. Yeganegi, S. R. Huisman, A. Lagendijk, and W. L. Vos, 
\emph{Inhibited spontaneous emission of quantum dots observed in a 3D photonic band gap},   
Phys. Rev. Lett. \textbf{107}, 193903 (2011).

\bibitem{Novotny2006book}
L. Novotny and B. Hecht, 
\emph{Principles of Nano-Optics }
(Cambridge University Press, Cambridge, 2006) 2nd Ed.

\bibitem{Soukoulis2011NP}
C.M. Soukoulis and M. Wegener, 
\emph{Past achievements and future challenges in the development of three-dimensional photonic metamaterials}, 
Nature Photonics {\bf 5}, 523 (2011) 

\bibitem{vandenBroek2012AFM}
J. M. van den Broek, L. A. Woldering, R. W. Tjerkstra, F. B. Segerink, I. D. Setija, and W. L. Vos, 
\emph{Inverse-woodpile photonic band gap crystals with a cubic diamond-like structure made from single-crystalline silicon},
Adv. Func. Mat. \textbf{22}, 25 (2012).

\bibitem{Woldering2009JAP}
L. A. Woldering, A. P. Mosk,  R. W. Tjerkstra, and W. L. Vos, 
\emph{The influence of fabrication deviations on the photonic band gap of three-dimensional inverse woodpile nanostructures},
J. Appl. Phys. \textbf{105}, 093108 (2009).

\end{thebibliography}

\section{Acknowledgments}
It is a great pleasure to salute Ad Lagendijk for his upcoming 70th birthday, and congratulate him with all seminal work he has done (and hopefully will continue to do!) in nanophotonics. 
It is a pleasure to thank Shanhui Fan (Stanford), Hasan Yilmaz (Yale), Alexandra Joita-Pacureanu (ESRF), Mehdi Aas for various help and discussions, and Hannie van den Broek, Leon Woldering, and Willem Tjerkstra for expert sample preparation, and the MESA+ cleanroom staff and ESRF staff for support. 
This work was supported by the ''Stirring of light!'' program of the "Stichting voor Fundamenteel Onderzoek der Materie" (FOM), which is financially supported by the "Nederlandse Organisatie voor Wetenschappelijk Onderzoek" (NWO), and also by the "Stichting voor Technische Wetenschappen" (STW).

\end{document}